\documentclass[12pt]{article}
\usepackage[cp1251]{inputenc}
\usepackage[russian]{babel}

\title{Канонический формализм релятивистской теории гравитации}
\author{В.О. Соловьев, М.В. Чичикина}
\date{}
\begin{document}
\maketitle
\begin{abstract}
Построен гамильтониан для релятивистской теории гравитации (РТГ) с
ненулевой массой гравитона. В качестве примера источника
гравитации рассматривается скалярное поле. Исключены связи второго
рода и построены скобки Дирака. Связи первого рода в теории
отсутствуют. Показано, что соответствующим образом выбирая
входящие в гамильтониан произвольные функции, можно получить
генераторы группы Пуанкаре. Их скобки Дирака реализуют алгебру
группы, в согласии с тем, что в РТГ имеются 10 законов сохранения.
\end{abstract}

\section{Введение}
Большинство физиков считает, что гравитационное поле, подобно остальным фундаментальным полям,
должно быть проквантовано. Хорошо известны трудности, с которыми сталкивается квантование поля метрического
тензора в общей теории относительности (ОТО). Представляет интерес вопрос о перспективах квантования
альтернативных теорий гравитации. Имея в виду, что каноническое квантование является исторически и не только
исторически, первым из методов объединения квантовой механики и классической теории, мы здесь, имея в виду
программу квантования релятивистской теории гравитации (РТГ)~\cite{Log}, начнем с построения для этой теории
гамильтонова формализма.  В раннем варианте~\cite{Sol86} решения задачи рассматривалась версия РТГ с нулевой
массой гравитона, от которой впоследствии отказались. Новая постановка, т.е присутствие ненулевой массы
гравитона, как будет видно из дальнейшего, приводит к существенно отличным результатам. В теории нет связей
первого рода, число степеней свободы возрастает,  инвариантность относительно группы Пуанкаре приводит к 10
интегралам движения.

Мы будем исходить из лагранжиана релятивистской теории гравитации~\cite{Log},
который позволяет более или менее стандартным способом осуществить
переход к гамильтониану и найти скобки Пуассона.

\section{Лагранжиан релятивистской теории гравитации}

В отличие от общей теории относительности релятивистская теория
гравитации содержит нединамическую плоскую метрику $h_{\mu\nu}$,
которая входит в лагранжиан теории наряду с динамической римановой
метрикой $g_{\mu\nu}$. Принимая скорость света равной единице мы
можем, согласно работе~\cite{Log},
 записать лагранжеву плотность гравитационного поля в виде
\begin{equation}
{\cal L}=\frac{1}{16\pi G}\sqrt{-g}R -\frac{m^2}{16\pi G}\left(\frac{1}{2}
h_{\mu\nu}\tilde g^{\mu\nu}-\sqrt{-g}-\sqrt{-h}\right) +\dots, \label{eq:Lagr}
\end{equation}
где многоточие обозначает поверхностные члены (4-дивергенции),
греческие индексы принимают
значения от 0 до 3, $G$ -- гравитационная постоянная, $g=\det(g_{\mu\nu})$,
$h=\det(h_{\mu\nu})$, $R$ -- скалярная кривизна пространства-времени,
определенная метрикой $g_{\mu\nu}$, $m$
 -- масса гравитона. Используется сигнатура $(-1,1,1,1)$.

С точностью до 4-дивергенций можно переписать лагранжеву плотность
(\ref{eq:Lagr}) в эквивалентном виде
\begin{equation}
{\cal L}=\frac{1}{16\pi G}
\tilde g^{\mu\nu}\left(\Delta\Gamma^{\lambda}_{\mu\sigma}\Delta\Gamma^{\sigma}
_{\nu\lambda}-\Delta\Gamma^{\lambda}_{\mu\nu}\Delta\Gamma^{\sigma}
_{\lambda\sigma}\right)
-\frac{m^2}{16\pi G}\left(\frac{1}{2}
h_{\mu\nu}\tilde g^{\mu\nu}-\sqrt{-g}-\sqrt{-h}\right) +\dots, \label{eq:Lagr2}
\end{equation}
где
\begin{equation}
\Delta\Gamma^{\lambda}_{\mu\nu}\equiv \bar\Gamma^{\lambda}_{\mu\nu}-
\Gamma^{\lambda}_{\mu\nu}
=\frac{1}{2}g^{\lambda\sigma}\left(
D_{\mu}g_{\sigma\nu}+D_{\nu}g_{\sigma\mu}-D_{\sigma}g_{\mu\nu}
\right),
\end{equation}
$\bar\Gamma^{\lambda}_{\mu\nu}$ -- символы Кристоффеля римановой метрики,
$\Gamma^{\lambda}_{\mu\nu}$ -- символы Кристоффеля плоской метрики,
а $D_{\mu}$ -- ковариантная производная, согласованная с плоской метрикой.

Независимыми переменными, подлежащими варьированию в действии, построенном
из лагранжевых плотностей (\ref{eq:Lagr}) или (\ref{eq:Lagr2}),
могут служить, например, 10 компонент римановой метрики $g_{\mu\nu}$.
Для упрощения формальных выкладок будут полезны также тензорные величины
\begin{equation}
f^{\mu\nu}\equiv\frac{\sqrt{-g}}{\sqrt{-h}}g^{\mu\nu}.
\end{equation}

\section{3+1-разложение тензоров и новое представление для лагранжевой
плотности}

При построении гамильтонова формализма необходимо выделить
направление эволюции, т.е. физическое время. При этом совсем
необязательно нарушать общую ковариантность теории и выбирать в
качестве времени одну из координат $X^{\alpha}$.  В работах
Кухаржа~\cite{Kuchar} было впервые показано, как следует строить
4-ковариантный канонический формализм. Здесь мы будем следовать
этому методу, пользуясь также результатами работы~\cite{Sol88}.

Фиксированному моменту физического времени соответствует некоторая пространственноподобная
гиперповерхность
\begin{equation}
X^{\alpha}=e^{\alpha}(x^i),
\end{equation}
где $x^i$ -- независимые координаты на гиперповерхности, латинские
индексы принимают значения от 1 до 3. В отличие от случаев,
рассмотренных в работах~\cite{Kuchar}, у нас имеется две метрики
пространства-времени, а не одна, и мы требуем, чтобы поверхность
была пространственноподобной по отношению к обеим
метрикам.\footnote{Возможность такого требования обеспечивается в
РТГ постулатом причинности.} Это означает наложение двух условий,
справедливых во всех точках гиперповерхности,
\begin{equation}
\gamma_{ij}(x^k)dx^idx^j>0,\quad \eta_{ij}(x^k)dx^idx^j>0,
\end{equation}
 где использованы две различные индуцированные метрики:
\begin{equation}
\gamma_{ij}=g_{\mu\nu}e^{\mu}_{,i}e^{\nu}_{,j},\quad
\eta_{ij}=h_{\mu\nu}e^{\mu}_{,i}e^{\nu}_{,j}.
\end{equation}
Очевидно, что в общем случае метрика $\eta_{ij}$, в отличие от $h_{\mu\nu}$,
не является плоской.

Далее, мы будем предполагать, что все пространство-время может быть заполнено
такими слоями постоянного физического времени, т.е. представлено в виде
однопараметрического семейства пространственноподобных гиперповерхностей:
\begin{equation}
X^{\alpha}=e^{\alpha}(x^i,t),
\end{equation}
причем можно ввести векторное поле
\begin{equation}
N^{\alpha}=\frac{\partial e^{\alpha}}{\partial t},
\end{equation}
которое будет всюду времениподобным по отношению к обеим метрикам
пространства-времени
\begin{equation}
g_{\alpha\beta}N^{\alpha}N^{\beta}<0,\quad
h_{\alpha\beta}N^{\alpha}N^{\beta}<0.
\end{equation}

Для $3+1$-разложения пространственно-временных тензоров необходимо ввести
базис, связанный с гиперповерхностью фиксированного времени. Мы могли бы,
например, использовать для этой цели четверку пространственно-временных
векторов $(N^{\alpha},e^{\alpha}_{,i})$. Однако на самом деле требуется
два базиса: один, связанный с метрикой $g_{\mu\nu}$, и второй, связанный
с метрикой $h_{\mu\nu}$. Раздвоение здесь происходит при переходе к нижним
индексам
\begin{equation}
N_{\beta}=h_{\beta\alpha}N^{\alpha},\quad
\bar N_{\beta}=g_{\beta\alpha}N^{\alpha},
\end{equation}
и аналогично, появляются $e_{\alpha i}$ и $\bar e_{\alpha i}$.
Технически более удобно ввести два других базиса, где в качестве
времениподобной составляющей выбираются векторы единичной нормали к
гиперповерхности $(n^{\alpha},e^{\alpha}_{,i})$ и
$(\bar n^{\alpha},e^{\alpha}_{,i})$, определяемые, очевидно, соотношениями
\begin{equation}
h_{\alpha\beta}n^{\alpha}e^{\beta}_{,i}=0,\quad
h_{\alpha\beta}n^{\alpha}n^{\beta}=-1,
\end{equation}
\begin{equation}
g_{\alpha\beta}\bar n^{\alpha}e^{\beta}_{,i}=0,\quad
g_{\alpha\beta}\bar n^{\alpha}\bar n^{\beta}=-1.
\end{equation}

Теперь можно применять 3+1-разложение к величинам различной тензорной
размерности, например,
\begin{eqnarray}
N^{\alpha}&=&Nn^{\alpha}+N^ie^{\alpha}_{,i}=\bar N\bar n^{\alpha}+\bar N^i
e^{\alpha}_{,i},\nonumber\\
g^{\mu\nu}&=&g^{\perp\perp}n^{\mu}n^{\nu}+g^{\perp j}n^{\mu}e^{\nu}_{,j}+
g^{i\perp}e^{\mu}_{,i}n^{\nu}
+g^{ij}e^{\mu}_{,i}e^{\nu}_{,j}=
(-1)\bar n^{\mu}\bar n^{\nu}
+\gamma^{ij}e^{\mu}_{,i}e^{\nu}_{,j},\nonumber\\
f^{\mu\nu}&=&f^{\perp\perp}n^{\mu}n^{\nu}+f^{\perp j}n^{\mu}e^{\nu}_{,j}+
f^{i\perp}e^{\mu}_{,i}n^{\nu}
+f^{ij}e^{\mu}_{,i}e^{\nu}_{,j}
,\label{eq:3+1}
\end{eqnarray}
где
\begin{equation}
N=-n_{\mu}N^{\mu},\quad N^i=e_{\mu}^{i}N^{\mu},\quad \bar N=
-\bar n_{\mu}N^{\mu},
\quad \bar N^i=\bar e_{\mu}^iN^{\mu},\label{eq:parameters}
\end{equation}
\begin{equation}
g^{\perp\perp}=n_{\mu}n_{\nu}g^{\mu\nu},\quad g^{\perp j}=
g^{j\perp}=-n_{\mu}e_{\nu}^j
g^{\mu\nu},
\quad g^{ij}=e_{\mu}^ie_{\nu}^jg^{\mu\nu},\dots
\end{equation}
Нетрудно установить линейную связь векторов двух базисов
\begin{equation}
\bar n^{\alpha}=\sqrt{-g^{\perp\perp}}n^{\alpha}-\frac{g^{i\perp}}{\sqrt{
-g^{\perp\perp}}}e^{\alpha}_i,
\end{equation}
и, соответственно, линейную связь составляющих, например,
\begin{equation}
\bar N=-\frac{1}{f^{\perp\perp}}\sqrt{\frac{\gamma}{\eta}}N,\quad
\bar N^i=N^i-\frac{f^{\perp i}}{f^{\perp\perp}}N.\label{eq:NN}
\end{equation}

\section{Построение канонического формализма для гравитационного поля}

Для преобразования к нужному виду плотности лагранжиана (\ref{eq:Lagr})
необходимо преобразовать два слагаемых, первое из которых не содержит ни
массы гравитона, ни плоской метрики. Поэтому для первого слагаемого можно
воспользоваться стандартными преобразованиями~\cite{Kuchar} ОТО, после чего, с точностью до поверхностных членов, оно
оказывается равным
\begin{equation}
-\frac{N}{16\pi G}
\frac{\gamma}{f^{\perp\perp}\sqrt{\eta}}(\tilde R-{\bar K}^2+{\rm Sp}{\bar K}
^2),
\end{equation}
где $\tilde R$
  -- скалярная кривизна гиперповерхности, построенная с помощью метрики
$\gamma_{ij}$,   $\bar K_{ij}$
-- вторая фундаментальная форма гиперповерхности в римановой
геометрии, заданной метрикой  $g_{\mu\nu}$,  ${\rm Sp}{\bar K}^2=
\bar K_{ij}\bar K_{kl}\gamma^{ik}\gamma^{jl}$.
 Второе слагаемое, с учетом формулы
\begin{equation}
f^{ij}=\frac{1}{f^{\perp\perp}}\left(
f^{\perp i}f^{\perp j}-\frac{\gamma\gamma^{ij}}{\eta}
\right),
\end{equation}
и после подстановки в него разложений
(\ref{eq:3+1}),          принимает вид
\begin{equation}
-N\sqrt{\eta}\frac{m^2}{16\pi G}\left[-1-\frac{f^{\perp\perp}}{2}+
\frac{f^{\perp i}f^{\perp j}\eta_{ij}}{2f^{\perp\perp}}-
\frac{1}{f^{\perp\perp}}\frac{\gamma}{\eta}\left(\frac{1}{2}
\eta_{ij}\gamma^{ij}-1
\right)
\right],
\end{equation}
как видно, оно не содержит скоростей и поэтому не влияет на определения
импульсов.

Поверхностные вклады в лагранжеву плотность (полные производные по
времени и пространственные дивергенции) не влияют на
симплектическую структуру, и следовательно, на скобки Пуассона,
граничные условия на пространственной бесконечности принимаются
такими, что риманова метрика стремится к плоской метрике
Минковского, а гиперповерхности стремятся к гиперплоскостям.
Поэтому приходим к действию для гравитационного поля вида
\begin{eqnarray}
S&=&
\int\limits^{t_2}_{t_1}dt\int\limits_{R^3}d^3x
\left(
-\frac{N}{16\pi G}\frac{\gamma}{f^{\perp\perp}\sqrt{\eta}}
(\tilde R-{\bar K}^2+{\rm Sp}{\bar K}^2)
\right.
\nonumber\\
&-&
\left.
N\sqrt{\eta}\frac{m^2}{16\pi G}
\left[-1-\frac{f^{\perp\perp}}{2}+
\frac{f^{\perp i}f^{\perp j}\eta_{ij}}{2f^{\perp\perp}}-
\frac{1}{f^{\perp\perp}}\frac{\gamma}{\eta}
\left(\frac{1}{2}\eta_{ij}\gamma^{ij}-1\right)
\right]
\right).\label{eq:action}
\end{eqnarray}
При этом независимыми переменными, подлежащими варьированию, мы считаем
$\gamma_{ij}(x^k,t)$, $f^{\perp\perp}(x^k,t)$, $f^{\perp i}(x^k,t)$.
Известными и поэтому не подлежащими варьированию следует считать плоскую
метрику пространства-времени $h_{\mu\nu}(X^{\alpha})$
и функции, задающие однопараметрическое
семейство пространственноподобных гиперповерхностей $e^{\alpha}(x^i,t)$,
через них в свою очередь выражаются векторы базиса $n^{\alpha}(x^i,t),
e^{\alpha}_i(x^i,t)$ и вектор $N^{\alpha}(x^i,t)$. Величины
$\bar K_{ij}(x^i,t)$ даются известными из канонического формализма
общей теории относительности формулами
\begin{equation}
\bar K_{ij}=\frac{1}{2\bar N}\left(\bar N_{i|j}+\bar N_{j|i}-\gamma_{ij,0},
\right)
\end{equation}
где в свою очередь функции $\bar N$ и $\bar N^i$ выражаются через $N$, $N^i$
и $f^{\perp\perp}$, $f^{\perp i}$ формулами (\ref{eq:NN}).
Вертикальная черта обозначает ковариантную производную в римановой геометрии
3-мерного пространства,
определяемую метрикой $\gamma_{ij}$.
Чтобы не спутать обозначения импульсов с отношением длины окружности к ее
диаметру, спрячем последнее в новую константу $\kappa=16\pi G$.

Исходя из действия (\ref{eq:action})
для сопряженных импульсов находим соотношения
\begin{eqnarray}
\pi_{\perp}&=&\frac{\partial{\cal L}}{\partial f^{\perp\perp}_{,0}}=0,
\label{eq:constraint1}\\
\pi_{i}&=&\frac{\partial{\cal L}}{\partial f^{\perp i}_{,0}}=0,
\label{eq:constraint2}\\
\pi_{ij}&=&\frac{\partial{\cal L}}{\partial\gamma_{ij,0}}=
\frac{\partial{\cal L}}{\partial\bar K_{ij}}\frac{\partial\bar K_{ij}}
{\partial\gamma_{ij,0}}=-\frac{\sqrt{\gamma}}{\kappa}
(\bar K^{ij}-\gamma^{ij}
\bar K),
\end{eqnarray}
из которых видно, что (\ref{eq:constraint1}) и  (\ref{eq:constraint2})
являются первичными связями в терминологии Дирака~\cite{Dirac}
и должны быть добавлены к гамильтониану теории с произвольными множителями
Лагранжа. Таким образом, получаем
\begin{equation}
{\rm H}=\int\limits_{R^3}d^3x\left(
\pi^{ij}\gamma_{ij,0}-{\cal L} +\lambda^{\perp}\pi_{\perp}+\lambda^i\pi_i
\right),
\end{equation}
где необходимо выразить скорости через импульсы по формуле
\begin{equation}
\gamma_{ij,0}=\bar N_{i|j}+\bar N_{j_i}+\frac{2\kappa\bar N}{\sqrt{\gamma}}
(\pi_{ij}-\gamma_{ij}\frac{\pi}{2}).
\end{equation}
После этой процедуры гамильтониан гравитационного поля,
с точностью до поверхностных членов, принимает вид
\begin{equation}
{\rm H}=\int\limits_{R^3}d^3x\left(
N{\cal H}+N^i{\cal H}_i +\lambda^{\perp}\pi_{\perp}+\lambda^i\pi_i
\right),\label{eq:Ham1}
\end{equation}
где
\begin{eqnarray}
{\cal H}&=&-\frac{1}{f^{\perp\perp}}\sqrt{\frac{\gamma}{\eta}}{\bar{\cal H}}-
\frac{f^{\perp i}}{f^{\perp\perp}}{\bar{\cal H}}_i\nonumber\\
&+&\frac{m^2\sqrt{\eta}}{\kappa}\left[
-1-\frac{f^{\perp\perp}}{2}+
\frac{f^{\perp i}f^{\perp j}\eta_{ij}}{2f^{\perp\perp}}-
\frac{1}{f^{\perp\perp}}\frac{\gamma}{\eta}\left(\frac{1}{2}
\eta_{ij}\gamma^{ij}-1
\right)
\right],\\
{\cal H}_i&=&\bar{\cal H}_i=-2\pi_{i|j}^j,\\
\bar{\cal H}&=&-\frac{1}{\sqrt{\gamma}}\left(
\frac{1}{\kappa}\gamma\tilde R+\kappa(\frac{\pi^2}{2}-\mathrm{Sp}\pi^2)
\right).
\end{eqnarray}

Канонические скобки Пуассона
\begin{equation}
\{F,G\}=\int\limits_{R^3}d^3x\left[\frac{\delta F}{\delta\gamma_{ij}}
\frac{\delta G}{\delta\pi^{ij}}+
\frac{\delta F}{\delta f^{\perp\perp}}
\frac{\delta G}{\delta\pi_{\perp}}
+
\frac{\delta F}{\delta f^{\perp i}}
\frac{\delta G}{\delta\pi_{i}}-(F\leftrightarrow G)
\right]\label{eq:PB}
\end{equation}
позволяют записать гамильтоновы уравнения в привычном виде
\begin{equation}
\gamma_{ij,0}=\{\gamma_{ij}, {\rm H}\},\quad
\pi^{ij}_{,0}=\{\pi^{ij}, {\rm H}\},\label{eq:he1}
\end{equation}
\begin{equation}
f^{\perp\perp}_{,0}=\{f^{\perp\perp},{\rm H}\},\quad
\pi_{\perp,0}=\{\pi_{\perp}, {\rm H}\},\label{eq:he2}
\end{equation}
\begin{equation}
f^{\perp i}_{,0}=\{f^{\perp i},{\rm H}\},\quad
\pi_{i,0}=\{\pi_{i}, {\rm H}\}.\label{eq:he3}
\end{equation}

Далее необходимо убедиться, что первичные связи (\ref{eq:constraint1}),
(\ref{eq:constraint2}) согласованы с уравнениями движения, для этого
следует обеспечить обращение в нуль производных по времени
$\pi_{\perp,0}$ и $\pi_{i,0}$.
Поскольку сопряженные переменные $f^{\perp\perp}$ и $f^{\perp i}$ входят
в гамильтониан алгебраически, мы получаем вторичные связи в виде
алгебраических уравнений
\begin{equation}
\frac{\partial{\cal H}}{\partial f^{\perp\perp}}=0,\quad
\frac{\partial{\cal H}}{\partial f^{\perp
i}}=0,\label{eq:secondconstraints}
\end{equation}
которые элементарно разрешаются и дают
\begin{eqnarray}
f^{\perp i}&=&\frac{\kappa}{m^2\sqrt{\eta}}\eta^{ij}\bar{\cal H}_j,
\label{eq:f_i}\\
f^{\perp\perp}&=&-\frac{\kappa}{m^2\sqrt{\eta}}\sqrt{
\eta^{ij}\bar{\cal H}_i\bar{\cal H}_j+2\frac{m^2\sqrt{\eta}}{\kappa}
\left[
\sqrt{\frac{\gamma}{\eta}}\bar{\cal H}+\frac{m^2\sqrt{\eta}}{\kappa}
\frac{\gamma}{\eta}\left(\frac{1}{2}\eta_{ij}\gamma^{ij}-1\right)
\right]\label{eq:f_perp}
}.
\end{eqnarray}
Из разрешенного вида вторичных связей легко увидеть, что их скобки Пуассона
с первичными связями отличны от нуля, т.е. все связи являются связями второго
рода и могут быть полностью исключены введением скобок Дирака.
В данном случае скобки Дирака получаются из скобок Пуассона (\ref{eq:PB})
простым исключением членов с переменными $(f^{\perp\perp},\pi_{\perp})$  и
$(f^{\perp i},\pi_i)$
\begin{equation}
\{F,G\}_D=\int\limits_{R^3}d^3x\left[\frac{\delta F}{\delta\gamma_{ij}}
\frac{\delta G}{\delta\pi^{ij}}
-\frac{\delta F}{\delta\pi^{ij}}\frac{\delta G}{\delta\gamma_{ij}}
\right].\label{eq:DB}
\end{equation}
Подставляя решения уравнений связи в гамильтониан получаем
\begin{eqnarray}
{\rm H}&=&\int\limits_{R^3}d^3x
\left[
N\left(
\sqrt{
\eta^{ij}\bar{\cal H}_i\bar{\cal H}_j+2\frac{m^2\sqrt{\eta}}{\kappa}
\left[
\sqrt{\frac{\gamma}{\eta}}\bar{\cal H}+\frac{m^2\sqrt{\eta}}{\kappa}
\frac{\gamma}{\eta}
\left(
\frac{1}{2}\eta_{ij}\gamma^{ij}-1
\right)
\right]}\right.\right.\nonumber\\
&-&\left.\left.\frac{m^2\sqrt{\eta}}{\kappa}
\right)
+N^i\bar{\cal H}_i
\right].\label{eq:Ham2}
\end{eqnarray}

Мы пришли к гамильтониану, который зависит от канонических переменных
$\gamma_{ij},\pi^{ij}$, а также содержит зависимость от известной заранее
(определяемой из фиксированной метрики пространства-времени
и функций, определяющих гиперповерхности) метрики $\eta_{ij}$.
Входящая константа обеспечивает нормировку энергии вакуума:
если риманова метрика совпадает с плоской $g_{\mu\nu}=h_{\mu\nu}$,
то при любом задании гиперповерхностей получаем ${\rm H}=0$.

Добавим, что мы могли бы не сразу исключать из
гамильтониана все связи и
вводить скобки Дирака, а сначала согласовать вторичные связи с динамикой
(\ref{eq:he1}) -- (\ref{eq:he2}) -- (\ref{eq:he3}), нетрудно видеть, что
это дало бы нам возможность найти определение лагранжевых множителей
через канонические переменные, но не привело бы к новым связям.
После подстановки найденных лагранжевых множителей в уравнения
(\ref{eq:he2}) -- (\ref{eq:he3}) мы получили бы соотношения, эквивалентные
так называемым условиям гармоничности
\begin{equation}
D_{\mu}f^{\mu\nu}=0.
\end{equation}
Однако, введение скобок Дирака исключает самостоятельную роль этих
уравнений и они становятся следствием гамильтоновых уравнений
движения, порожденных гамильтонианом (\ref{eq:Ham2}) и скобками (\ref{eq:DB}).

\section{Скалярное поле как пример источника гравитации}

Разумеется, наш формализм будет неполным без демонстрации подключения
материальных полей. Предположим, что взаимодействие этих полей с гравитацией
является минимальным, тогда их
плотность лагранжиана ${\cal L}_M$
будет зависеть от набора полей $\phi^A(X^{\alpha})$, их первых
производных по координатам пространства-времени и от метрики
$g_{\mu\nu}(X^{\alpha})$, преобразуясь как скалярная плотность при общих
координатных преобразованиях. Приведение к гамильтонову виду действия
полей материи выполняется согласно процедуре Кухаржа~\cite{Kuchar} и
в результате дает
\begin{equation}
S_M=\int\limits^{t_2}_{t_1}dt\int\limits_{R^3}d^3x\left(\pi_A\phi^A_{,0}-
\bar N\bar{\cal H}_M-\bar N^i\bar{\cal H}_{Mi}\right).\label{eq:maction}
\end{equation}
Объединение этого действия с действием гравитации (\ref{eq:action})
сводится к тому, что $\bar{\cal H}_M$  и $\bar{\cal H}_{Mi}$ просто
добавляются к тем $\bar{\cal H}$  и $\bar{\cal H}_i$, которые раньше
содержали только переменные гравитационного поля.

Для иллюстрации рассмотрим скалярное поле с плотностью лагранжиана
\begin{equation}
{\cal L}_M=-\sqrt{-g}\left(
\frac{1}{2}g^{\mu\nu}\partial_{\mu}\phi\partial_{\nu}\phi+U(\phi)
\right)
\end{equation}
Переход к 3+1-обозначениям и преобразование Лежандра могут быть выполнены
для скалярного поля независимо  от чисто гравитационного вклада
\begin{equation}
{\cal L}_M=-N\sqrt{\eta}\left(f^{\perp\perp}\phi_{,\perp}\phi_{,\perp}
+2f^{\perp i}\phi_{,\perp}\phi_{,i}+\frac{1}{f^{\perp\perp}}\left(
f^{\perp i}f^{\perp j}-\frac{\gamma\gamma^{ij}}{\eta}
\right)\phi_{,i}\phi_{,j}+U(\phi)
\right),
\end{equation}
причем
\begin{equation}
\phi_{,0}=-N\phi_{,\perp}+N^i\phi_{,i}.
\end{equation}
Импульс определяется обычным образом
\begin{equation}
\pi_{\phi}=\frac{\partial {\cal L}}{\partial\phi_{,0}}
=\sqrt{\eta}\left(f^{\perp\perp}\phi_{,\perp}-f^{\perp i}\phi_{,i}
\right),
\end{equation}
а скорость выражается через него по формуле
\begin{equation}
\phi_{,0}=-\frac{N}{f^{\perp\perp}\sqrt{\eta}}\pi_{\phi}+\left(N^i-
N\frac{f^{\perp i}}{f^{\perp\perp}}\right)\phi_{,i}.
\end{equation}
После соответствующего преобразования Лежандра действие скалярного поля
принимает вид (\ref{eq:maction}) где
\begin{equation}
\bar{\cal H}_M=\frac{1}{\sqrt{\gamma}}\left(
\frac{\pi^2_{\phi}}{2}+\frac{1}{2}\gamma\gamma^{ij}\partial_i\phi\partial_j\phi
+\gamma U(\phi)
\right),\quad
\bar{\cal H}_{Mi}=\pi_{\phi}\phi_{,i}.
\end{equation}
Таким образом, объединяя действия для гравитационного и скалярного
полей мы получим те же самые первичные связи
(\ref{eq:constraint1}), (\ref{eq:constraint2}), что и в случае
чистой гравитации, а полный гамильтониан, содержащий связи, будет
иметь тот же вид (\ref{eq:Ham1}). Процедура исключения связей
также не меняется и окончательный гамильтониан сохраняет вид
(\ref{eq:Ham2}), причем теперь
\begin{eqnarray}
\bar{\cal H}_i&=&-2\pi_{i|j}^j+\pi_{\phi}\phi_{,i},\nonumber\\
\bar{\cal H}&=&\frac{1}{\sqrt{\gamma}}\left(-
\frac{1}{\kappa}\gamma\tilde R+\kappa(\mathrm{Sp}\pi^2
-\frac{\pi^2}{2}) +
\frac{\pi^2_{\phi}}{2}+\frac{1}{2}\gamma\gamma^{ij}\partial_i\phi\partial_j\phi
+\gamma U(\phi) \right),\nonumber\\
\{F,G\}_D&=&\int\limits_{R^3}d^3x\left[\frac{\delta
F}{\delta\gamma_{ij}} \frac{\delta G}{\delta\pi^{ij}}+\frac{\delta
F}{\delta\phi} \frac{\delta G}{\delta\pi_\phi} -\frac{\delta
F}{\delta\pi^{ij}}\frac{\delta G}{\delta\gamma_{ij}} -\frac{\delta
F}{\delta\pi_\phi}\frac{\delta
G}{\delta\phi}\right].\label{eq:DB2}
\end{eqnarray}
Нетрудно убедиться, что гамильтоновы уравнения движения для
системы взаимодействующих скалярного и гравитационного полей имеют
вид (для случая  $U(\phi)=1/2M^2\phi^2$):
\begin{eqnarray}
\phi_{,0}&=&\int\limits_{R^3}d^3x\Bigl(
\bar N\{\phi,\bar{\cal H}\}_D+\bar N^k\{\phi,\bar{\cal H}_k\}_D
\Bigr)\nonumber\\
&=&\bar N\frac{\pi_\phi}{\sqrt{\gamma}}+{\bar N}^i\phi_{,i},\\
\pi_{\phi,0}&=&\int\limits_{R^3}d^3x\Bigl(
\bar N\{\pi_{\phi},\bar{\cal H}\}_D+\bar N^k\{\pi_{\phi},\bar{\cal H}_k\}_D
\Bigr)\nonumber\\
&=&(\bar N\sqrt{\gamma}\gamma^{ij}\partial_j\phi)_{,i}-\bar N\sqrt{\gamma}M^2
\phi+(\bar N^i\pi_\phi)_{,i},\\
\gamma_{ij,0}&=&\int\limits_{R^3}d^3x\Bigl(
\bar N\{\gamma_{ij},\bar{\cal H}\}_D+\bar N^k\{\gamma_{ij},\bar{\cal H}_k\}_D
\Bigr)\nonumber\\
&=&{\bar N}_{i|j}+{\bar N}_{j|i}+\kappa\frac{2\bar N}{\sqrt{\gamma}}(\pi_{ij}
-\gamma_{ij}\frac{\pi}{2}),\\
\pi^{ij}_{,0}&=&\int\limits_{R^3}d^3x\Bigl(
\{\pi^{ij},\bar N\bar{\cal H}\}_D+\bar N^k\{\pi^{ij},\bar{\cal H}_k\}_D
\Bigr)\nonumber\\
&+&\frac{m^2}{\kappa}\bar N\sqrt{\gamma}\left[
\gamma^{ij}+\frac{1}{2}\eta_{kl}\left(
\gamma^{ki}\gamma^{lj}-\gamma^{ij}\gamma^{kl}
\right)
\right]\nonumber\\
&=& -\frac{1}{2}\bar N\sqrt{\gamma}(\gamma^{ij}\gamma^{mn}-\gamma^{im}
\gamma^{jn})\partial_m\phi\partial_n\phi-\frac{1}{2}\bar N\sqrt{\gamma}
\gamma^{ij}M^2\phi^2\nonumber\\
&-&\frac{1}{\kappa}\bar N\sqrt{\gamma}(R^{ij}-\gamma^{ij}R)+\kappa\frac{\bar
N}{\sqrt{\gamma}}(\pi\pi^{ij}-2\pi^{ik}\pi^j_k)\nonumber\\
&+&\frac{1}{\kappa}
\sqrt{\gamma}({\bar N}^{|ij}-\gamma^{ij}\bar N^{|k}_{|k})+
(\pi^{ij}{\bar N}^k)_{|k}-\pi^{ik}{\bar N}^j_{|k}-\pi^{kj}{\bar
N}^i_{|k}\nonumber\\
&+&\frac{m^2}{\kappa}\bar N\sqrt{\gamma}\left[
\gamma^{ij}+\frac{1}{2}\eta_{kl}\left(
\gamma^{ki}\gamma^{lj}-\gamma^{ij}\gamma^{kl} \right) \right],
\end{eqnarray}
из которого очевидно, что отличие от соответствующих уравнений ОТО
проявляется только в последнем уравнении и имеет порядок величины
$O(m^2/\kappa)$. Зависимость величин $f^{\perp\perp}$, $f^{\perp
i}$ от основных переменных (\ref{eq:f_perp}), для которых выше
были получены гамильтоновы уравнения, можно не учитывать при
вычислении скобок Дирака в силу формул
(\ref{eq:secondconstraints}).

\section{Группа Пуанкаре в гамильтоновом формализме РТГ}
Среди всех вариантов гамильтоновой эволюции, разнообразие которых проистекает из произвола в выборе функций $N(x)$, $N^i(x)$ в гамильтониане~(\ref{eq:Ham2}), содержатся преобразования, сохраняющие метрику Минковского. Выбирая в качестве гиперповерхностей гиперплоскости и выбирая на них декартовы координаты, мы получаем на гиперплоскостях метрику $\eta_{ij}$, индуцированную метрикой Минковского (7), в простейшем виде $\eta_{ij}=\delta_{ij}$, а функции преобразований~(\ref{eq:parameters})  в виде
\begin{equation}
N=A_kx^k+a,\quad N^i=A_{ik}x^k+a^i,\label{eq:Poincare}
\end{equation}
где
\[A_{ik}=-A_{ki}.\]
Тогда гамильтониан~(\ref{eq:Ham2}), ввиду его линейности по функциям $N(x)$, $N^i(x)$, примет вид
\begin{equation}
H=P^0a-P^ia^i+M^kA_k+\frac{1}{2}M^{ik}A_{ik},
\end{equation}
где
\begin{eqnarray}
P^0 &=&-\frac{m^2}{\kappa}\int \left(1+f^{\perp\perp}\right)d^3x, \\
P_i &=& -\frac{m^2}{\kappa}\int f^{\perp i}d^3x\equiv-\int{\cal H}_i d^3x, \\
M^{ik} &=& -\frac{m^2}{\kappa} \int\left(x^i f^{\perp k}-x^k f^{\perp i}\right)d^3x\equiv\int\left(x^k{\cal H}_i-x^i{\cal H}_k\right)d^3x, \\
M^k &=& -\frac{m^2}{\kappa}\int x^k(1+f^{\perp\perp})d^3x.
\end{eqnarray}
Смысл этих операторов явствует из того, что все они являются частными случаями гамильтониана, соответствующими различному выбору преобразований координат, ими генерируемых: $P^0$ отвечает преобразованию сдвига по времени, $P^i$ -- пространственным трансляциям, $M^{ik}$ -- пространственным поворотам и $M^k$ -- бустам.
Наши обозначения выбраны для удобства сравнения с аналогичными
формулами работы~\cite{RT}, где рассматривалась алгебра Пуанкаре в
асимптотически плоском пространстве ОТО. 

Однако прежде, чем
сводить гамильтониан к такому упрощенному виду, полезно получить
алгебру скобок Дирака (\ref{eq:DB2}) для общих гамильтонианов. Пусть
\begin{eqnarray}
{\rm H}&=&\int\limits_{R^3}d^3x\Biggl(N{\cal H}+N^i{\cal
H}_i\nonumber
\\&=&\int\limits_{R^3}d^3x\Biggl(
\bar N\bar {\cal H}+\bar N^i\bar{\cal H}_i\nonumber\\
&+&\frac{m^2\sqrt{\eta}}{\kappa}N\left[
-1-\frac{f^{\perp\perp}}{2}+ \frac{f^{\perp i}f^{\perp
j}\eta_{ij}}{2f^{\perp\perp}}-
\frac{1}{f^{\perp\perp}}\frac{\gamma}{\eta}\left(\frac{1}{2}
\eta_{ij}\gamma^{ij}-1 \right) \right]\Biggr),\label{eq:Ham3}
\end{eqnarray}
где мы считаем $f^{\perp\perp}$, $f^{\perp i}$ функциями, имеющими
нулевые скобки Дирака. Это оправдано тем, что связи второго рода
можно учитывать как до, так и после вычисления скобок Дирака.
Таким образом, первая половина связей (\ref{eq:constraint1}),
(\ref{eq:constraint2}) учитывается до, а вторая половина
(\ref{eq:f_i}), (\ref{eq:f_perp}) -- после. Как обычно, при
расчетах отбрасываются все поверхностные интегралы. Это оправдано для островных систем, где излучение допускается только во внутренней области, но не на бесконечности, а скорость стремления римановой метрики к плоской определяется юкавским поведением.

Результаты вычислений можно представить либо в виде, удобном для
сравнения с аналогичной формулой ОТО:
\begin{eqnarray}
\{H(\alpha, \alpha^i),H(\beta,\beta^j)\}&=&\int d^3x \Bigl[\bar\lambda\bar{\cal H}+\bar\lambda^k\bar{\cal H}_k+(\bar\alpha\bar\beta^k_{|k}-\bar\beta\bar\alpha^k_{|k})\bar{\cal H}\nonumber\\&-&\frac{m^2}{\kappa}\sqrt{\gamma}\gamma_{k\ell}(\bar\alpha\bar\beta^{k|\ell}-\bar\beta\bar\alpha^{k|\ell})
(2-\eta_{mn}\gamma^{mn}))\nonumber\\
&-&\frac{m^2}{\kappa}\sqrt{\gamma}\eta_{k\ell}(\bar\alpha\bar\beta^{k|\ell}-\bar\beta\bar\alpha^{k|\ell})\label{eq:algebra_g}
\Bigr],\\
\bar\lambda&=&\bar\alpha^i\bar\beta_{,i}-\bar\beta^i\bar\alpha_{,i},\\
\bar\lambda^k&=&\gamma^{k\ell}(\bar\alpha\bar\beta_{,\ell}-\bar\beta\bar\alpha_{,\ell})+\bar\alpha^\ell\bar\beta^k_{,\ell}-\bar\beta^\ell\bar\alpha^k_{,\ell},\label{eq:algebra_par}
\end{eqnarray}
либо в виде, соответствующем теориям поля на фоне фиксированной
метрики:
\begin{eqnarray}
\{H(\alpha, \alpha^i),H(\beta,\beta^j)\}&=& H(\lambda,
\lambda^k)+\int \frac{\partial {\cal H}}{\partial \eta_{ij}}
\left(\alpha{\cal L}_{\vec\beta}\eta_{ij}-\beta{\cal
L}_{\vec\alpha}\eta_{ij}\right)d^3x,\label{eq:algebra_h}\\
\lambda&=&\alpha^i\beta_{,i}-\beta^i\alpha_{,i},\\
\lambda^k&=&\eta^{k\ell}(\alpha\beta_{,\ell}-\beta\alpha_{,\ell})+\alpha^\ell\beta^k_{,\ell}-\beta^\ell\alpha^k_{,\ell},
\end{eqnarray}
где ${\cal L}_{\vec\alpha}\eta_{ij}$ -- производная Ли от метрики
$\eta_{ij}$ по направлению векторного поля  $\vec\alpha$.
 Отличия от
ОТО проявляются в (\ref{eq:algebra_g}) как в членах,
пропорциональных квадрату массы гравитона, так и в коэффициенте
при $\bar{\cal H}$. Последнее связано с тем, что коэффициент при
$\bar{\cal H}$, т.е. функция $\bar N$, пропорционален
$\sqrt{\gamma}$. Соотношения (\ref{eq:algebra_g}), таким образом,
не представляют собой известную алгебру деформаций
гиперповерхности~\cite{Dirac},\cite{RT}, т.к. функции $\bar N$,
$\bar N^i$ не являются ее параметрами.

Подстановка в соотношения~(\ref{eq:algebra_h}) вместо произвольных
функций $\alpha$, $\alpha^i$, $\beta$, $\beta^j$ выражений вида
(\ref{eq:Poincare}), отвечающих преобразованиям Пуанкаре, приводит
к соотношениям алгебры Пуанкаре для скобок Дирака:
\begin{eqnarray}
  \{P^0,P_i\}_D &=& 0, \quad  \{P_i,P_j\}_D = 0, \\
  \{P^0,M^{ik}\}_D &=& 0, \quad  \{P_i,M^{jk}\}_D = \delta_{ik}P_j-\delta_{ij}P_k, \\
  \{M^{ij},M^{k\ell}\}_D
&=&\delta_{ik}M^{j\ell}-\delta_{i\ell}M^{jk}+\delta_{j\ell}M^{ik}-\delta_{jk}M^{i\ell}, \\
  \{P^0,M^i\}_D &=& -P^i, \quad  \{P_i,M^j\}_D =  -\delta_{ij}(P^0-c^0),\\
 \{M^k,M^{ij}\}_D&=&\delta_{kj}(M^i-c^i)-\delta_{ki}(M^j-c^j),\quad \{M^i,M^j\}_D=-M^{ij}.
\end{eqnarray}
Аддитивные вклады $c^0=m^2/\kappa\int d^3x$ и $c^i=m^2/\kappa\int x^id^3x$ в $P_0$ и $M^i$, не зависящие от канонических
переменных и выражаюшиеся расходящимися интегралами по всему
пространству, играют роль центральных зарядов в канонической
реализации алгебры Пуанкаре и отвечают классической перенормировке
энергии вакуума. Они возникают вследствие желания обеспечить строго нулевую плотность энергии для пустого пространства Минковского. С этой целью в гамильтониан (и в лагранжиан) включается член нулевого порядка по физическому полю. В то же время линейные по физическому полю члены, как в лагранжиане, так и в гамильтониане, появляются только в виде полных производных и не дают вклада в уравнения движения.

\section{Заключение}

Попытаемся теперь резюмировать, в чем состоит сходство и в чем различие
между РТГ и ОТО при формулировке на языке гамильтонова формализма.

Мы видим, что канонические переменные и их скобки в обеих теориях
совпадают, однако, гамильтонианы отличаются. Различие во внешнем
виде гамильтонианов оказывается, однако, не самым существенным и,
действуя формально, его можно свести к минимуму, переписав формулу
(\ref{eq:Ham2}) в виде (\ref{eq:Ham3}), куда требуется подставлять
выражения для $f^{\perp\perp},f^{\perp i}$ из соотношений
(\ref{eq:f_i}), (\ref{eq:f_perp}) после вычисления скобок Дирака.
Более важным является то обстоятельство, что одни и те же величины
$\bar{\cal H},\bar{\cal H}_i$  в ОТО должны обращаться в нуль на
любых решениях уравнений движения, а в РТГ это требование
отсутствует.

Это различие ведет к тому, что число степеней свободы в теориях не
совпадает. В РТГ мы  имеем 6 чисто гравитационных степеней свободы
на точку пространства, не считая обычного числа степеней свободы
полей материи (для скалярного поля это число, очевидно, равно
1). В ОТО из числа 6 мы должны вычесть количество связей первого
рода, $6-4=2$, т.е. получаем только две степени свободы.

Отсутствие связей первого рода приводит к тому, что гамильтонианы
РТГ, в частности, генераторы группы Пуанкаре, не сводятся к поверхностным интегралам на
решениях уравнений движения, в отличие от гамильтонианов ОТО.  Таким образом, в рамках РТГ определена плотность энергии-импульса и других интегралов движения.

Вопрос о знаке плотности энергии в полученном выражении (57) требует дополнительного изучения. В линеаризованном приближении, с точностью до пространственной дивергенции, плотность энергии является знакопеременной квадратичной формой. Однако существенное значение, конечно, имеет только полная теория. Если бы в этой теории существовала патология в виде отрицательного потока энергии от скалярной компоненты, то она проявлялась бы, например, в виде излучения сферических волн. Но в работе~\cite{LM} было показано, что излучение скалярной компоненты гравитационного поля в сферически симметричном случае отсутствует и внешнее поле является исключительно статическим.

Вопросы канонического квантования мы надеемся рассмотреть в следующих работах.

\end{document}